\documentclass[12pt,a4]{article}
\usepackage{sw20lart}

\input tcilatex
\QQQ{Language}{
American English
}

\begin{document}

\title{\textbf{Three dimensional fermionic determinants, Chern-Simons and nonlinear
field redefinitions }}
\author{\textbf{D.G. Barci, V.E.R. Lemes, C. Linhares de Jesus, } \and \textbf{%
M.B.D. Silva Maia Porto, S.P. Sorella, L.C.Q.Vilar} \and \vspace{2mm} \\
UERJ, Universidade do Estado do Rio de Janeiro\\
Departamento de F\'\i sica Te\'orica\\
Instituto de F\'{\i}sica\\
Rua S\~ao Francisco Xavier, 524\\
20550-013, Maracan\~{a}, Rio de Janeiro, Brazil \\
\vspace{2mm} \and \textbf{UERJ/DFT-01/98}\vspace{2mm}\newline \and \textbf{%
PACS: 11.10.Gh}}
\maketitle

\begin{abstract}
The three dimensional abelian fermionic determinant of a two component
massive spinor in flat euclidean space-time is resetted to a pure
Chern-Simons action through a nonlinear redefinition of the gauge field.

\setcounter{page}{0}\thispagestyle{empty}
\end{abstract}

\vfill\newpage\ \makeatother
\renewcommand{\theequation}{\thesection.\arabic{equation}}

\section{\ Introduction\-}

Recently, it has been shown \cite{lett1,lett2} that any local Yang-Mills
type action\footnote{%
According to the BRST analysis of gauge theories \cite{bdk,dv,bbh}, the name
Yang-Mills type action is employed here to denote a generic integrated local
invariant polynomial built only with the field strength $F$ and its
covariant derivatives.} in the presence of the topological three dimensional
Chern-Simons term can be reabsorbed into the pure Chern-Simons through a
local covariant nonlinear gauge field redefinition. Choosing in fact as the
Yang-Mills action the standard $\int FF\;$term, we have \cite{lett1,lett2}

\begin{equation}
\mathcal{S}_{CS}(A)+\frac 1{4m}tr\int d^3xF_{\mu \nu }F^{\mu \nu }=\mathcal{S%
}_{CS}(\widehat{A})\;,  \label{reab}
\end{equation}
with

\begin{equation}
\widehat{A}_\mu =A_\mu +\sum_{k=1}^\infty \frac 1{m^k}\vartheta _\mu ^k\;,
\label{nl}
\end{equation}
and

\begin{equation}
\mathcal{S}_{CS}(A)=\frac 12tr\int d^3x\varepsilon ^{\mu \nu \rho }\left(
A_\mu \partial _\nu A_\rho +\frac 23gA_\mu A_\nu A_\rho \right) \;.
\label{cs}
\end{equation}
The two parameters $g,m$ in the expressions $\left( \text{\ref{reab}}\right)
-\left( \text{\ref{cs}}\right) $ identify respectively the gauge coupling
and the so called topological mass \cite{tmym,pr}. The coefficients $%
\vartheta _\mu ^k$ in the eq.$\left( \text{\ref{nl}}\right) $ turn out to be
local\textit{\ }and covariant, meaning that they are built only with the
field strength $F_{\mu \nu }$

\begin{equation}
F_{\mu \nu }=\partial _\mu A_\nu -\partial _\nu A_\mu +g[A_\mu ,A_\nu ]\;,
\label{fs}
\end{equation}
and its covariant derivative $D_\mu $

\begin{equation}
D_\mu =\partial _\mu +g\left[ A_\mu ,\;\right] \;.  \label{cov-der}
\end{equation}
For instance, the first four terms of the expansion $\left( \text{\ref{nl}}%
\right) $ have been found to be \cite{lett1}

\begin{eqnarray}
\vartheta _\mu ^1 &=&\frac 14\varepsilon _{\mu \sigma \tau }F^{\sigma \tau
}\;,  \label{coeff} \\
\vartheta _\mu ^2 &=&\frac 18D^\sigma F_{\sigma \mu }\;,  \nonumber \\
\vartheta _\mu ^3 &=&-\frac 1{16}\varepsilon _{\mu \sigma \tau }D^\sigma
D_\rho F^{\rho \tau }\;+\frac g{48}\varepsilon _{\mu \sigma \tau }\left[
F^{\sigma \rho },F_\rho ^{\;\tau }\right] \;,  \nonumber \\
\vartheta _\mu ^4 &=&-\frac 5{128}D^2D^\rho F_{\rho \mu }+\frac 5{128}D^\nu
D_\mu D^\lambda F_{\lambda \nu }\;  \nonumber \\
&&-\frac 7{192}g\left[ D^\rho F_{\rho \tau },F_\mu ^{\;\;\tau }\right]
\;-\frac 1{48}g\left[ D_\nu F_{\mu \lambda },F^{\lambda \nu \;}\right] \;. 
\nonumber
\end{eqnarray}
The formulas $\left( \text{\ref{reab}}\right) ,\left( \text{\ref{nl}}\right)
\;$can be generalized to any higher dimensional Yang-Mills term built with
the field strength $F$ and its covariant derivatives, expressing therefore
the classical equivalence, up to nonlinear field redefinitions, among the
Yang-Mills actions in the presence of the Chern-Simons and the pure
Chern-Simons term.

Although this equivalence has been rigorously proven \cite{lett1,lett2} only
for the class of the local Yang-Mills type terms, it has been suggested that
it could persist at the level of the $1PI$ effective quantum action. More
precisely, it has been argued \cite{lett2} that the complete $1PI$ effective
action obtained upon quantization in the Landau gauge of the massive
Yang-Mills action $\left( \text{\ref{reab}}\right) $ could be cast in the
form of a pure Chern-Simons through a nonlinear and nonlocal gauge field
redefinition. This hypothesis has been tested on a class of nonlocal gauge
invariant terms expected to contribute to the $1PI$ effective action of
topological massive Yang-Mills \cite{lett2}. As one can easily understand,
the nonlocality of the field redefinition in the quantum case stems from the
fact that the loop corrections to the effective $1PI$ action give rise to
both local and nonlocal gauge invariant terms.

The aim of this work is to provide a further evidence in favour of this
hypothesis by means of a direct example of a three dimensional system whose
corresponding quantum effective action can be fully resetted to pure
Chern-Simons, up to a nonlocal nonlinear redefinition of the gauge
connection. The model we will refer to is the abelian fermionic determinant
of a massive two component spinor interacting with an external gauge field
in flat euclidean space-time \cite{tmym,ns,red,cl}. In particular, we shall
be able to prove that the infinite number of one loop diagrams representing
the perturbative expansion of the fermionic determinant can be reabsorbed
into pure Chern-Simons, up to field redefinitions. Remarkably in half and in
spite of its nonlocal character, the redefined field turns out to transform
still as a connection. This property will be of great relevance in order to
give a geometrical interpretation of the final result.

It is worth recalling here that the fermionic determinant plays a rather
important role in different areas of theoretical physics, going from pure
solid state applications \cite{frad} to the three dimensional bosonization 
\cite{ma,kk,fs1,bu,fs2,marc,barci,mb}$.$

The present work is organized as follows. In Sect.2 we present the strategy
which will be adopted in order to reset the fermionic determinant to pure
Chern-Simons. In Sect.3 we establish a very useful cohomological recursive
formula. Sect.4 is devoted to the analysis of the fermionic determinant.
Sect.5 deals with the generalization to a family of determinants with
nonminimal gauge interaction. Finally, in Sect.6 we collect the concluding
remarks and we outline possible further applications.

\section{The strategy}

The effective action $\Gamma (A)$ generated by a two component massive
spinor interacting with an abelian external gauge field can be written, via
functional integral, as

\begin{eqnarray}
e^{\Gamma (A)} &=&\int \mathcal{D}\psi \mathcal{D}\overline{\psi }\;e^{\int
d^3x\overline{\psi }(i\gamma ^\mu \partial _\mu +\gamma ^\mu A_\mu -m)\psi
}\;,  \label{det} \\
\Gamma (A) &=&\text{\textrm{logdet}}(i\gamma ^\mu \partial _\mu +\gamma ^\mu
A_\mu -m)\;.  \nonumber
\end{eqnarray}
Being the gauge field $A_\mu \;$an external field, the perturbative
expansion of $\Gamma (A)$ consists of an infinite series of one loop
diagrams. The strategy which shall be adopted in order to reset the
expression $\left( \text{\ref{det}}\right) $ to the Chern-Simons action
relies on the analysis of the exact expression available for the two point
function \cite{tmym,ns,red,cl,gam,rothe,cc} and on two observations
concerning the structure of a generic $n$-point, $n\geq 2,$ contribution to
the effective action $\Gamma (A)$.

As it is well known, the one loop two point function (\textit{i.e. }the
spinor vacuum polarization) has been computed exactly. Although a detailed
discussion will be given in Sect.4, it is worth emphasizing here that the
complete contribution of the vacuum polarization contains several terms,
among which the abelian Chern-Simons action

\begin{equation}
\mathcal{S}_{CS}^{ab}(A)=\frac 12\int d^3x\varepsilon ^{\mu \nu \rho }A_\mu
\partial _\nu A_\rho \;.  \label{ab-cs}
\end{equation}
Of course, the presence of this term will be crucial for the purposes of the
present work.

Concerning now the aforementioned observations, the first one aims at
showing that any action of the type

\begin{equation}
\mathcal{S}_\Omega (A)=\frac 1\xi \int d^3x_1....d^3x_nF_{\mu _1\nu
_1}(x_1)....F_{\mu _n\nu _n}(x_n)\Omega ^{\mu _1\nu _1...\mu _n\nu
_n}(x_1,...,x_n)\;,  \label{om-act}
\end{equation}
with $n\geq 2$ and $\Omega ^{\mu _1\nu _1...\mu _n\nu _n}(x_1,...,x_n)\;$%
being a generic (\textit{even nonlocal}) space-time dependent kernel%
\footnote{%
Although the Lorentz structure of the kernel $\Omega ^{\mu _1\nu _1...\mu
_n\nu _n}$ can be specified by means of suitable combinations of the flat
euclidean metric and of the $\varepsilon_{\mu \nu \rho}$ tensor, we leave it
unspecified for the sake of generality.}, can be reabsorbed into the pure
Chern-Simons action $\left( \text{\ref{ab-cs}}\right) $ through a nonlinear
redefinition, namely

\begin{equation}
\mathcal{S}_{CS}^{ab}(A)+\mathcal{S}_\Omega (A)=\mathcal{S}_{CS}^{ab}(%
\widehat{A})\;,  \label{om-red}
\end{equation}
with

\begin{equation}
\widehat{A}_\mu =A_\mu +\sum_{k=1}^\infty \frac 1{\xi ^k}\vartheta _\mu ^k\;.
\label{xi-red}
\end{equation}
The parameter $\xi $ in eq.$\left( \text{\ref{om-act}}\right) $ is an
arbitrary coefficient. As it will be discussed in the next Section, this
step will be handled by means of a useful recursive cohomological formula.
In addition, it will be checked that the redefined field $\widehat{A}_\mu \;$%
turns out to transform as a connection.

The second observation concerns the higher order terms of the effective
action $\Gamma (A)$. We shall be able to prove that the contributions with $%
n\geq 2$ external gauge fields can be cast in the form of the eq.$\left( 
\text{\ref{om-act}}\right) $, due to the absence of anomalies in three
dimensions. This result combined with the knowledge of the spinor vacuum
polarization will enable us to reset the fermionic determinant to pure
Chern-Simons.

\section{A recursive formula}

The task of this section is to establish a simple recursive formula
accounting for the eqs.$\left( \text{\ref{om-red}}\right) ,\left( \text{\ref
{xi-red}}\right) $. For a better understanding of the mechanism which allows
us to reabsorb the term $\mathcal{S}_\Omega (A)$ into the pure Chern-Simons,
let us first work out some of the coefficients $\vartheta _\mu ^k$ of the
expansion $\left( \text{\ref{xi-red}}\right) $. Their computation is rather
straightforward. After inserting the eq.$\left( \text{\ref{xi-red}}\right) $
in the eq.$\left( \text{\ref{om-red}}\right) $ and identifying the terms
with the same power in the inverse of the parameter $\xi $, for the
coefficients $\vartheta _\mu ^1,$ $\vartheta _\mu ^2,$ $\vartheta _\mu
^3,\vartheta _\mu ^4$ we obtain

\begin{equation}
\vartheta _\mu ^1(x)=\varepsilon _{\mu \nu \rho }\Xi ^{\nu \rho }(x)\;,
\label{t1}
\end{equation}
\begin{eqnarray}
\vartheta _\mu ^2(x) &=&-\varepsilon _{\mu \nu \rho }\int d^3y\mathcal{F}%
_{3n}^{\sigma \tau \nu \rho }(y,x)\varepsilon _{\sigma \tau \alpha }\partial
_\beta ^y\Xi ^{\alpha \beta }(y)\;  \nonumber  \label{t2} \\
\mathcal{F}_{3n}^{\sigma \tau \nu \rho } &=&\int \left(
\prod_{j=3}^nd^3x_jF_{\mu _j\nu _j}(x_j)\right) \Omega ^{\sigma \tau \nu
\rho \mu _3\nu _3..\mu _n\nu _n}(y,x,x_3,..,x_n)\;,  \nonumber  \label{t2} \\
&&  \label{t2}
\end{eqnarray}
\begin{eqnarray}
\frac 12\vartheta _\mu ^3(x) &=&\varepsilon _{\mu \nu \rho }\int d^3yd^3z%
\mathcal{F}_{4n}^{\sigma \tau \lambda \delta \nu \rho }(y,z,x)\varepsilon
_{\sigma \tau \alpha }\partial _\beta ^y\Xi ^{\alpha \beta }(y)\varepsilon
_{\lambda \delta \omega }\partial _\tau ^z\Xi ^{\omega \tau }(z)\;  \nonumber
\label{t3} \\
\mathcal{F}_{4n}^{\sigma \tau \lambda \delta \nu \rho } &=&\int \left(
\prod_{j=4}^nd^3x_jF_{\mu _j\nu _j}(x_j)\right) \Omega ^{\sigma \tau \lambda
\delta \nu \rho \mu _4\nu _4..\mu _n\nu _n}(y,z,x,x_4,..,x_n)\;,  \nonumber
\label{t3} \\
&&  \label{t3}
\end{eqnarray}

\begin{eqnarray}
-\frac 14\vartheta _\mu ^4(x) &=&\varepsilon _{\mu \nu \rho }\int
d^3td^3yd^3z\mathcal{F}_{5n}^{\sigma \tau \lambda \delta \alpha \beta \nu
\rho }\varepsilon _{\sigma \tau \gamma }\partial _\tau ^z\Xi _z^{\gamma \tau
}\varepsilon _{\lambda \delta \omega }\partial _\eta ^y\Xi _y^{\omega \eta
}\varepsilon _{\alpha \beta \xi }\partial _\theta ^t\Xi _t^{\xi \theta } 
\nonumber  \label{t4} \\
&&-\frac 18\varepsilon _{\mu \nu \rho }\int d^3yd^3z\mathcal{F}_{4n}^{\alpha
\beta \xi \chi \nu \rho }\varepsilon _{\alpha \beta \sigma }\partial _\tau
^y\Xi _y^{\sigma \tau }\varepsilon _{\xi \chi \eta }\varepsilon ^{\omega
\eta \lambda }\partial _\lambda ^z\vartheta _\omega ^2(z)\;,  \nonumber \\
\mathcal{F}_{5n}^{\sigma \tau \lambda \delta \alpha \beta \nu \rho } &=&\int
\left( \prod_{j=5}^nd^3x_jF_{\mu _j\nu _j}(x_j)\right) \Omega ^{\sigma \tau
\lambda \delta \alpha \beta \nu \rho \mu _5\nu _5..\mu _n\nu
_n}(t,y,z,x,x_5,..,x_n)\;,  \nonumber  \label{t4} \\
&&  \label{t4}
\end{eqnarray}
with

\begin{equation}
\Xi ^{\nu \rho }(x)=\int d^3x_2....d^3x_nF_{\mu _2\nu _2}(x_2)....F_{\mu
_n\nu _n}(x_n)\Omega ^{\nu \rho \mu _2\nu _2...\mu _n\nu
_n}(x,x_2,...,x_n)\;.  \label{xi-coeff}
\end{equation}
Observe that, as already mentioned, the expressions in eqs.$\left( \text{\ref
{t1}}\right) $-$\left( \text{\ref{t4}}\right) $ are gauge invariant. As a
consequence, the redefined field $\widehat{A}_\mu \;$behaves as a connection
under gauge transformations

\begin{equation}
\delta A_\mu (x)=-\partial _\mu \alpha (x)\;,\;\;\;\;\;\;\;\;\;\delta 
\widehat{A}_\mu (x)=-\partial _\mu \alpha (x)\;.  \label{g-tr}
\end{equation}
It follows then that the resulting Chern-Simons action $\mathcal{S}%
_{CS}^{ab}(\widehat{A})\;$is gauge invariant as well, \textit{i.e.}

\begin{equation}
\delta \mathcal{S}_{CS}^{ab}(\widehat{A})=0\;.  \label{inv-ab-cs}
\end{equation}
Although the higher order coefficients of the expansion $\left( \text{\ref
{xi-red}}\right) $ can be easily obtained, let us present here a
cohomological recursive argument for the equation $\left( \text{\ref{om-red}}%
\right) $. To this purpose, we introduce a ghost field $c$ and a set of
antifields $A_\mu ^{*},c^{*}$ in order to implement in cohomology the
equations of motion stemming from the action

\begin{equation}
\mathcal{S}_{CS}^{ab}(A)+\mathcal{S}_\Omega (A)\;.  \label{in-act}
\end{equation}
For the BRST differential we have

\begin{eqnarray}
sA_\mu &=&-\partial _\mu c\;,  \label{brst} \\
sc &=&0\;,  \nonumber \\
sA_\mu ^{*} &=&\frac{\delta (\mathcal{S}_{CS}^{ab}+\mathcal{S}_\Omega )}{%
\delta A^\mu }\;=\frac 12\varepsilon _{\mu \nu \rho }F^{\nu \rho }+\frac{%
\delta \mathcal{S}_\Omega (A)}{\delta A^\mu }\;,  \nonumber \\
sc^{*} &=&-\partial ^\mu A_\mu ^{*}\;,  \nonumber
\end{eqnarray}
where

\begin{eqnarray}
\frac{\delta \mathcal{S}_\Omega (A)}{\delta A^\mu } &=&\;\frac 2\xi \int
\left( \prod_{j=2}^nd^3x_jF_{\mu _j\nu _j}(x_j)\right) \partial _\nu
^x\Omega ^{\mu \nu \mu _2\nu _2...\mu _n\nu _n}(x,x_2,..,x_n)\;  \label{eq-m}
\\
&&+\frac 2\xi \int \left( \prod_{j\neq 2}^nd^3x_jF_{\mu _j\nu
_j}(x_j)\right) \partial _\nu ^x\Omega ^{\mu _1\nu _1\mu \nu ...\mu _n\nu
_n}(x_1,x,x_3..,x_n)  \nonumber \\
&&+\;...........  \nonumber \\
&&+\frac 2\xi \int \left( \prod_{j=1}^{n-1}d^3x_jF_{\mu _j\nu
_j}(x_j)\right) \partial _\nu ^x\Omega ^{\mu _1\nu _1...\mu _{n-1}\nu
_{n-1}\mu \nu }(x_1,..,x_{n-1},x)\;.  \nonumber
\end{eqnarray}
The fields and antifields $A_\mu ,c,A_\mu ^{*},c^{*}$ possess respectively
ghost number $0$, $1$, $-1$, $-2$. Following now refs.\cite{lett1,lett2}, it
is easily established that the third equation of $\left( \text{\ref{brst}}%
\right) $ can be cast in the form of a recursive formula. In fact,
contracting both sides with $\varepsilon _{\mu \nu \rho }$ and using

\begin{equation}
\varepsilon _{\mu \nu \rho }\varepsilon ^{\rho \sigma \tau }=\delta _\mu
^\sigma \delta _\nu ^\tau -\delta _\nu ^\sigma \delta _\mu ^\tau \;,
\label{norm}
\end{equation}
we get

\begin{eqnarray}
F_{\mu \nu } &=&s(\varepsilon _{\mu \nu \rho }A^{*\rho })-\varepsilon _{\mu
\nu \rho }\frac{\delta \mathcal{S}_\Omega (A)}{\delta A_\rho }\;
\label{rec-f} \\
&=&s(\varepsilon _{\mu \nu \rho }A^{*\rho })-\frac{2\varepsilon _{\mu \nu
\rho }}\xi \int \left( \prod_{j=2}^nd^3x_jF_{\mu _j\nu _j}(x_j)\right)
\partial _\tau ^x\Omega ^{\rho \tau \mu _2\nu _2...\mu _n\nu
_n}(x,x_2,..,x_n)  \nonumber \\
&&\;-\frac{2\varepsilon _{\mu \nu \rho }}\xi \int \left( \prod_{j\neq
2}^nd^3x_jF_{\mu _j\nu _j}(x_j)\right) \partial _\tau ^x\Omega ^{\mu _1\nu
_1\rho \tau ...\mu _n\nu _n}(x_1,x,x_3..,x_n)  \nonumber \\
&&\;-\;...........  \nonumber \\
&&\;-\frac{2\varepsilon _{\mu \nu \rho }}\xi \int \left(
\prod_{j=1}^{n-1}d^3x_jF_{\mu _j\nu _j}(x_j)\right) \partial _\tau ^x\Omega
^{\mu _1\nu _1...\mu _{n-1}\nu _{n-1}\rho \tau }(x_1,..,x_{n-1},x)\;. 
\nonumber
\end{eqnarray}
This equation has the meaning of an iterative formula since the field
strength $F_{\mu \nu }$ appears on both sides. At each step of the iteration
the $F_{\mu \nu }$ 's contained in the term $\delta \mathcal{S}_\Omega
/\delta A_\rho $ can be replaced by the exact BRST variation $s(\varepsilon
_{\mu \nu \rho }A^{\rho *})$ with the addition of terms of higher order in $%
1/\xi $. Obviously, the whole iteration procedure will result in a BRST
exact power series of the kind

\begin{equation}
F_{\mu \nu }=s\left( \varepsilon _{\mu \nu \rho }A^{*\rho
}+\sum_{k=1}^\infty \frac 1{\xi ^k}\mathcal{M}_{\mu \nu }^k(\Omega
,A,A^{*})\right) \;,  \label{f-ex}
\end{equation}
where the coefficients $\mathcal{M}_{\mu \nu }^k(\Omega ,A,A^{*})\;$depend
on the space-time kernel $\Omega $, the gauge field $A$, the antifield $%
A^{*} $ and their space-time derivatives. The formula $\left( \text{\ref
{f-ex}}\right) $ expresses the exactness of the field strength. Therefore we
can write the action $\mathcal{S}_\Omega (A)$ in the form of an exact
cocycle, \textit{i.e. }

\begin{eqnarray}
\mathcal{S}_\Omega  &=&\frac 1\xi s\int d^3x_1\prod_{j=2}^nd^3x_jF_{\mu
_j\nu _j}(x_j)\left( \varepsilon _{\mu _1\nu _1\rho }A^{*\rho
}+\sum_{k=1}^\infty \frac{\mathcal{M}_{\mu _1\nu _1}^k}{\xi ^k}\right)
\Omega ^{\mu _1\nu _1..\mu _n\nu _n}\;.  \nonumber  \label{ex-s-om} \\
&&  \label{ex--om-ac}
\end{eqnarray}
In turn, this implies that the action $\mathcal{S}_\Omega (A)$ can be
reabsorbed into pure Chern-Simons through a nonlinear field redefinition,
accounting then for the eq.$\left( \text{\ref{om-red}}\right) $. In fact,
from the eqs.$\left( \ref{brst}\right) $ it follows that the BRST variation
of antifield dependent expressions, as for instance $\left( \ref{ex--om-ac}%
\right) $, gives rise to terms which are proportional to the equations of
motion, thereby corresponding to field redefinitions. It is worth
underlining here that the possibility of reabsorbing the term $\mathcal{S}%
_\Omega (A)\;$depends crucially on the presence of the Chern-Simons in the
starting action. The recursive formula $\left( \text{\ref{rec-f}}\right) $
works in fact due to the presence of the field strength $F_{\mu \nu }$ in
the left hand side. Needless to say, this term follows from the field
variation of the Chern-Simons action $\mathcal{S}_{CS}^{ab}(A)$.

Let us conclude this Section by remarking that the use of the antifields and
of the BRST differential do not have here the meaning of a quantization
procedure. We are not attempting to quantize the action $\left( \text{\ref
{in-act}}\right) $. This would be a very hard task, due to the highly
nonlocal character of $\mathcal{S}_\Omega (A).$ The gauge field $A_\mu $ is
always meant to be an external classical field. The introduction of the
antifields has to be seen as a useful device in order to exploit from a
cohomological point of view the consequences following from the classical
equations of motion, as for instance the recursive formula $\left( \text{\ref
{rec-f}}\right) $. This role of the BRST differential is well known, being
related to the so called characteristic cohomology \cite{c-co} and to the
Koszul-Tate differential \cite{bbh,c-co}. Notice also that the formulas $%
\left( \text{\ref{f-ex}}\right) ,\left( \text{\ref{ex--om-ac}}\right) $ have
been derived by means of direct strightforward manipulations, without
relying on any particular propery of the BRST\ differential $\left( \text{%
\ref{brst}}\right) $ or on the explicit knowledge of its cohomology. $.$

\section{The fermionic determinant}

We are now ready to analyse the perturbative expansion of the fermionic
determinant. To this purpose we recall that because of charge conjugation
invariance the Green functions with an odd number of external gauge fields
vanish and that, although superficially divergent, the spinor vacuum
polarization turns out to be finite \cite{tmym,ns,red,cl,gam,rothe,cc}. The
higher $n$-point Green functions are finite by power counting. It is also
worth mentioning that in the infinite mass limit $m\rightarrow \infty $ the
fermionic determinant reduces, modulo the well known regularization
ambiguity \cite{tmym,ns,red,cl,gam,rothe,cc}, to the pure Chern-Simons
action.

Let us begin by showing that, due to the absence of gauge anomalies in three
dimensions, the generic $n$-point contribution to $\Gamma (A)$ can be cast
in the form of the eq.$\left( \text{\ref{om-act}}\right) .$

\subsection{Absence of anomalies and structure of the perturbation theory}

It is a well established fact that in three dimensions there are no gauge
anomalies, so that the functional $\Gamma (A)$ of eq.$\left( \text{\ref{det}}%
\right) $ is gauge invariant

\begin{equation}
\partial _\mu \frac{\delta \Gamma (A)}{\delta A_\mu }=0\;.  \label{g-inv}
\end{equation}
This equation has useful consequences. Acting indeed on $\left( \text{\ref
{g-inv}}\right) $ with the test operator

\begin{equation}
\frac \delta {\delta A_{\mu _1}(x_1)}\;.....\;\frac \delta {\delta A_{\mu
_{n-1}}(x_{n-1})}\;,  \label{test}
\end{equation}
and setting $A_\mu $ to zero we get

\begin{equation}
\left. \partial _{\mu _1}\frac \delta {\delta A_{\mu _1}(x_1)}\;.....\;\frac
\delta {\delta A_{\mu _n}(x_n)}\;\Gamma (A)\right| _{A=0}\;=0\;,  \label{wi}
\end{equation}
expressing the conservation law for the fermionic current insertions

\begin{equation}
\partial _{x_1}^{\mu _1}<j_{\mu _1}(x_1).....j_{\mu _n}(x_n)>=0\;,
\label{c-l}
\end{equation}
$j_\mu (x)\;$denoting the spinor current

\begin{equation}
j_\mu (x)=\overline{\psi }\gamma _\mu \psi \;.  \label{curr}
\end{equation}
The eq.$\left( \text{\ref{c-l}}\right) $ implies that the functional $\Gamma
(A)$ depends only on the gauge invariant variable $A_\mu ^T$ (see also App.A)

\begin{equation}
A_\mu ^T=(g_{\mu \nu }-\frac{\partial _\mu \partial _\nu }{\partial ^2}%
)A^\nu =A_\mu (x)-\partial _\mu ^x\partial _\nu ^x\int d^3y\;\mathcal{G}%
(x-y)A^\nu (y)\;,  \label{at}
\end{equation}
where

\begin{equation}
\mathcal{G}(x-y)=-\frac 1{4\pi }\frac 1{\left| x-y\right| }\;,  \label{gxy}
\end{equation}
is the inverse of the three dimensional laplacian

\begin{equation}
\partial _x^2\mathcal{G}(x-y)=\delta ^3(x-y)\;.  \label{inv}
\end{equation}
The nonlocal variable $A_\mu ^T\;$corresponds to the pure transverse part of
the gauge field $A_\mu \;$according to the decomposition

\begin{equation}
A_\mu =A_\mu ^T+A_\mu ^L\;,  \label{dec}
\end{equation}

\begin{equation}
A_\mu ^L=\frac{\partial _\mu \partial _\nu }{\partial ^2}A^\nu \;.
\label{al}
\end{equation}
Of course,

\begin{equation}
\partial ^\mu A_\mu ^T=0\;,  \label{tr}
\end{equation}
and

\begin{equation}
\delta A_\mu ^T=-\partial _\mu \alpha +\partial _\mu \alpha =0\;.
\label{at-g}
\end{equation}
Let us consider then the perturbative expansion of $\Gamma (A)$

\begin{eqnarray}
\Gamma (A) &=&\sum_{n=2}^\infty \Gamma ^n(A)\;,  \label{pert} \\
\Gamma ^n(A) &=&\int d^3x_1....d^3x_nA^{\mu _1}(x_1)...A^{\mu
_n}(x_n)<j_{\mu _1}(x_1).....j_{\mu _n}(x_n)>\;.  \nonumber
\end{eqnarray}
It is almost immediate now to verify that we can replace $A_\mu $ by $A_\mu
^T$ in the eq.$\left( \text{\ref{pert}}\right) $. In fact, due to the
conservation law $\left( \text{\ref{c-l}}\right) $, the longitudinal
components $A_\mu ^L$ do not couple to the spinor current, \textit{i.e. }

\begin{eqnarray}
0 &=&-\int d^3x_1....d^3x_n\left( \frac{\partial _\nu }{\partial ^2}A^\nu
(x_1)\right) ...A^{\mu _n}(x_n)\partial _{x_1}^{\mu _1}<j_{\mu
_1}(x_1).....j_{\mu _n}(x_n)>\;  \nonumber  \label{lo} \\
&=&\int d^3x_1....d^3x_nA^{L\mu _1}(x_1)...A^{\mu _n}(x_n)<j_{\mu
_1}(x_1).....j_{\mu _n}(x_n)>\;.  \label{lo}
\end{eqnarray}
Thus

\begin{equation}
\Gamma ^n(A)=\Gamma ^n(A^T)\;.  \label{tr-pert}
\end{equation}
Recalling now that

\begin{eqnarray}
F_{\mu \nu } &=&\partial _\mu A_\nu -\partial _\nu A_\mu =\partial _\mu
A_\nu ^T-\partial _\nu A_\mu ^T\;,  \label{div} \\
\partial ^\mu F_{\mu \nu } &=&\partial ^2A_\nu ^T\;,
\end{eqnarray}
it follows

\begin{equation}
A_\nu ^T=\frac 1{\partial ^2}\partial ^\mu F_{\mu \nu }=\frac 1{4\pi }\int
d^3y\frac{(x-y)^\mu }{\left| x-y\right| ^3}F_{\mu \nu }(y)\;.  \label{nice}
\end{equation}
Finally, for the $n$-point contribution $\Gamma ^n(A)\;$we get

\begin{equation}
\Gamma ^n(A)=\int d^3y_1....d^3y_nF_{\mu _1\nu _1}(y_1)....F_{\mu _n\nu
_n}(y_n)\Omega ^{\mu _1\nu _1...\mu _n\nu _n}(y_1,...,y_n)\;,  \label{fin-np}
\end{equation}
with the space-time kernel $\Omega ^{\mu _1\nu _1...\mu _n\nu
_n}(y_1,...,y_n)$ given by

\begin{eqnarray}
\Omega ^{\mu _1\nu _1...\mu _n\nu _n} &=&\frac 1{\left( 4\pi \right) ^n}\int
\left( \prod_{j=1}^nd^3x_j\frac{(x_j-y_j)^{\mu _j}}{\left| x_j-y_j\right| ^3}%
\;\right) <j^{\nu _1}(x_1)...j^{\nu _n}(x_n)>\;,  \nonumber  \label{kern} \\
&&  \label{kern}
\end{eqnarray}
where a suitable antisymmetrization in the Lorentz indices $(\mu _j,\nu
_j)\; $has to be understood. We see then that, as announced, the $n$-point
contribution to the effective action $\Gamma ^n(A)$ can be cast in the form
of the eq.$\left( \text{\ref{om-act}}\right) .$ It remains now to analyse
the two point function. This will be the task of the next Subsection.

\subsection{The spinor vacuum polarization}

The conclusions of the previous Subsection can be generalized to the
fermionic determinant in higher space-time dimensions, provided one is able
to guaranty the absence of anomalies. However, the three dimensional case is
peculiar with respect to the higher dimensional ones. Of course, the
peculiarity lies in the appearence of the Chern-Simons action $\mathcal{S}%
_{CS}^{ab}(A)$ in the two point function, as it has been established by the
various exact computations of the one loop spinor vacuum polarization done
till now \cite{tmym,ns,red,cl,gam,rothe,cc}.

Although higher dimensional generalizations of the Chern-Simons are known,
it is only in three dimensions that the field variation of the Chern-Simons
action yields the field strength $F_{\mu \nu }$. As already underlined, it
is this property which allows us to reabsorb gauge invariant $F$-dependent
actions through nonlinear redefinitions.

Owing to the results \cite{tmym,ns,red,cl,gam,rothe,cc}, the contribution of
the spinor vacuum polarization to the effective action can be written as

\begin{equation}
\Gamma ^2(A)=\Gamma ^2(A^T)=\eta \mathcal{S}_{CS}^{ab}(A)+\Gamma _F^2(A)\;,
\label{v-s-p}
\end{equation}
where $\eta $ is the well known regularization ambiguity \cite
{tmym,ns,red,cl,gam,rothe,cc} and where $\Gamma _F^2(A)\;$has the general
form $\left( \text{\ref{fin-np}}\right) $. In particular it turns out that,
when expanded in the inverse of the mass parameter $m$, $\Gamma _F^2(A)$
leads to an infinite sum of terms of the type

\begin{equation}
\int d^3xF_{\mu \nu }(\partial ^2)^{n-1}F^{\mu \nu }\;,\;\;\;\;\;\;\;\;\int
d^3x\varepsilon ^{\mu \nu \rho }A_\mu \partial _\nu (\partial ^2)^nA_\rho
\;,\;\;\;\;\;\;n\geq 1\;.  \label{terms}
\end{equation}
Observe that

\begin{equation}
\int d^3x\varepsilon ^{\mu \nu \rho }A_\mu \partial _\nu (\partial
^2)^nA_\rho =-\int d^3x\varepsilon ^{\mu \nu \rho }F_{\mu \sigma }\partial
_\nu (\partial ^2)^{n-1}F_\rho ^{\;\sigma }\;,  \label{equiv}
\end{equation}
$\;$has indeed the form of the eq.$\left( \text{\ref{fin-np}}\right) $. $\;$

We also remark that the presence of the Chern-Simons term in the eq.$\left( 
\text{\ref{v-s-p}}\right) $ is in complete agreement with the general fact
that the effective action $\Gamma $ depends only on the transverse component 
$A_\mu ^T$. It is almost immediate to check that the Chern-Simons term is in
fact already purely transverse,

\begin{equation}
\int d^3x\varepsilon ^{\mu \nu \rho }A_\mu \partial _\nu A_\rho =\int
d^3x\varepsilon ^{\mu \nu \rho }A_\mu ^T\partial _\nu A_\rho ^T\;.
\label{cs-tr}
\end{equation}

\subsection{Resetting the fermionic determinant to pure Chern-Simons}

To summarize, we have been able to show that the perturbative expansion of
the fermionic determinant can be written as

\begin{equation}
\Gamma (A^T)=\eta \mathcal{S}_{CS}^{ab}(A)+\sum_{n\geq 2}\int
d^3y_1...d^3y_nF_{\mu _1\nu _1}...F_{\mu _n\nu _n}\Omega ^{\mu _1\nu
_1...\mu _n\nu _n}\;\;,  \label{f-sum}
\end{equation}
for a suitable space-time dependent kernel $\Omega ^{\mu _1\nu _1...\mu
_n\nu _n}.$ Therefore, owing to the results of Sect.3, we can write for the
whole quantum action $\Gamma $ the following formula

\begin{equation}
\Gamma (A)=\text{\textrm{logdet}}(i\gamma ^\mu \partial _\mu +\gamma ^\mu
A_\mu -m)=\eta \mathcal{S}_{CS}^{ab}(\widehat{A})\;,  \label{res}
\end{equation}
up to a nonlinear redefinition $\widehat{A}$ of the gauge field of the kind
of eq.$\left( \text{\ref{xi-red}}\right) .$

This formula is the essence of the present work, expressing the fact that
the quantum effects can be reabsorbed in the pure Chern-Simons action, up to
nonlinear field redefinitions.

Although being out of the aim of this work, let us emphasize that the
equation $\left( \text{\ref{res}}\right) $ calls for a deeper understanding
of the regularization ambiguity coefficient $\eta $ \cite
{tmym,ns,red,cl,gam,rothe,cc}.

\section{Generalization}

It is not difficult now to prove that a result similar to $\left( \text{\ref
{res}}\right) $ holds in the case in which the spinor fields interact in a
nonminimal way with the gauge field $A_\mu .$ For instance, the inclusion of
an interaction term of the type

\begin{equation}
\varepsilon ^{\mu \nu \rho }\overline{\psi }\gamma _\mu \psi F_{\nu \rho }\;,
\label{nm-i}
\end{equation}
will not alter the gauge invariance of the resulting fermionic determinant.
As a consequence, the $n$-point contribution to the effective action will be
always of the form $\left( \text{\ref{fin-np}}\right) .$

Therefore, up to nonlinear field redefinitions, we get

\begin{equation}
\text{\textrm{logdet}}(i\gamma ^\mu \partial _\mu +\gamma ^\mu A_\mu +\frac
g2\varepsilon ^{\mu \nu \rho }\gamma _\mu F_{\nu \rho }-m)=\eta ^{\prime }%
\mathcal{S}_{CS}^{ab}(\widehat{A})\;.  \label{gen}
\end{equation}
The above formula generalizes to any higher dimensional $F$-dependent
nonminimal interaction, implying   that a whole family of quantum effective
actions can be actually resetted to pure Chern-Simons, up to nonlinear field
redefinitions. The suggestive picture which emerges from these results is
that the introduction of a  nonminimal gauge coupling in the fermionic
determinant corresponds to a change of the redefined  connection $\widehat{A}%
_\mu $ for the resulting Chern-Simons. In other words, it seems rather
natural to interpret the Chern-Simons as a gauge invariant functional
defined on the space of the connections of the type $\left( \text{\ref
{xi-red}}\right) .$ One moves from a given determinant to another one by an
appropriate change of the connection $\widehat{A}_\mu $. This point will
lead to rather interesting conclusions.

\section{Conclusion}

Several remarks follow from the previous considerations.

\begin{itemize}
\item  The gauge invariance of the coefficients $\vartheta _\mu ^k\;$%
entering the nonlinear redefinition of the gauge field (see eqs.$\left( 
\text{\ref{t1}}\right) $-$\left( \text{\ref{t4}}\right) $) implies that the
redefined field $\widehat{A}_\mu $ is still a connection. As already
underlined in ref.\cite{lett2}, this feature allows us to interpret the
resulting Chern-Simons action $\mathcal{S}_{CS}^{ab}(\widehat{A})$ as a
gauge invariant functional defined on the space of the connections of the
type $\left( \text{\ref{xi-red}}\right) .$ This provides a simple
geometrical set up for the fermionic determinant for an arbitrary finite
nonvanishing value of the mass parameter $m$.

\item  The possibility of resetting a whole family of fermionic determinants
(see for instance eq.$\left( \text{\ref{gen}}\right) $) to pure Chern-Simons
suggests the existence of a kind of universal behaviour for the
corresponding effective quantum actions. The universality factor is
precisely the Chern-Simons functional. The effective action of a given
determinant is obtained thus by evaluating the Chern-Simons functional at a
suitably chosen gauge connection $\widehat{A}_\mu $.

\item  This universality character should persist for any model which can be
related in some way to the fermionic determinant, as it is the case for
instance of the three dimensional Thirring model \cite{fs2,mb}. This point
could be of great relevance for the three dimensional bosonization program 
\cite{ma,kk,fs1,bu,fs2,marc,barci,mb}.

\item  Although the form of the coefficients $\vartheta _\mu ^k$ entering
the nonlinear redefinition $\left( \text{\ref{xi-red}}\right) $ relies on
the explicit computation of the space-time kernel $\Omega $ in eq.$\left( 
\text{\ref{f-sum}}\right) ,$ we believe that the knowledge of the fact that
the quantum effects can be reabsorbed into pure Chern-Simons and that the
resulting field $\widehat{A}_\mu $ is a connection represents a nonempty
information. Perhaps, the pure geometrical interpretation of the final
result could help us in finding a kind of recursive procedure for the $%
\vartheta _\mu ^k$ 's. This would allow us to obtain exact bosonized
formulas \cite{prog}. We observe also that the geometrical interpretation of
the nonlinear redefinition of the gauge field naturally reminds us the well
known normal coordinates expansion of the general relativity and of the
nonlinear sigma model.

\item  Finally, it is worth underlining that the present results yield a
further evidence in favour of the conjectured quantum equivalence \cite
{lett2}, up to nonlinear field redefinitions, between the pure nonabelian
Chern-Simons and the fully $1PI$ effective action of topological massive
Yang-Mills.
\end{itemize}

\vspace{5mm}

{\Large \textbf{Acknowledgements}}

The Conselho\ Nacional de Pesquisa e Desenvolvimento, CNP$q$ Brazil, the
Faperj, Funda\c {c}\~{a}o de Amparo \`{a} Pesquisa do Estado do Rio de
Janeiro and the SR2-UERJ are gratefully acknowledged for financial support.

\vspace{5mm}

\appendix

\section{Appendix}

\subsection{A gauge invariant perturbative expansion}

The dependence of the effective action $\Gamma (A)$ in eq.$\left( \text{\ref
{det}}\right) $ from the transverse component $A_\mu ^T\;$can also be seen
as a consequence of the invariance of the functional measure under phase
transformations of the type

\begin{equation}
\psi ^{\prime }=e^{-i\tau (x)}\psi \;,\;\;\;\;\;\;\;\;\overline{\psi }%
^{\prime }=e^{i\tau (x)}\overline{\psi }\;\;,  \label{c-v}
\end{equation}
namely

\begin{equation}
\mathcal{D}\psi ^{\prime }\mathcal{D}\overline{\psi }^{\prime }=\mathcal{D}%
\psi \mathcal{D}\overline{\psi }\;\;.  \label{inv-m}
\end{equation}
As it is well known, this property is related to the absence of anomalies in
three dimensions. In particular, this implies that the longitudinal
components $A_\mu ^L\;$can be completely gauged away. Moving in fact from $%
(\psi ,\overline{\psi \;})\;$to a set of gauge invariant spinor variables $%
(\chi ,\overline{\chi })$

\begin{eqnarray}
\chi (x) &=&e^{-i\int d^3y\mathcal{G}(x-y)\partial A(y)}\;\psi (x)\;,
\label{g-inv-s} \\
\overline{\chi }(x) &=&e^{+i\int d^3y\mathcal{G}(x-y)\partial A(y)}\;%
\overline{\psi }(x)\;,  \nonumber
\end{eqnarray}
with $\mathcal{G}(x-y)$ given in eq.$\left( \text{\ref{gxy}}\right) ,$ we get

\begin{equation}
\mathcal{D}\psi \mathcal{D}\overline{\psi }\;=\mathcal{D}\chi \mathcal{D}%
\overline{\chi }\;,  \label{p-chi}
\end{equation}
so that

\begin{equation}
\int \mathcal{D}\psi \mathcal{D}\overline{\psi }\;e^{\int d^3x\overline{\psi 
}(i\gamma ^\mu \partial _\mu +\gamma ^\mu A_\mu -m)\psi }\;=\int \mathcal{D}%
\chi \mathcal{D}\overline{\chi }\;e^{\int d^3x\overline{\chi }(i\gamma ^\mu
\partial _\mu +\gamma ^\mu A_\mu ^T-m)\chi }\;\;.  \label{p-i}
\end{equation}
This equation implies thus

\begin{equation}
\Gamma (A)=\Gamma (A^T)\;.  \label{nice}
\end{equation}

\end{document}